\documentclass[prl,twocolumn,superscriptaddress,showpacs]{revtex4}
\usepackage{graphicx}
\usepackage{amssymb}
\usepackage{amsmath}
\usepackage{bm}
\usepackage{color}
\usepackage[dvipdfm]{hyperref}

\newcommand{\figref}[1]{Fig.~\ref{#1}}

\newcommand{\bitem}{\begin{itemize}}
\newcommand{\eitem}{\end{itemize}}
\newcommand{\benum}{\begin{enumerate}}
\newcommand{\eenum}{\end{enumerate}}
\newcommand{\btab}[1]{\begin{tabular}{#1}}
\newcommand{\etab}{\end{tabular}}
\newcommand{\btabn}[1]{\begin{tabular}{#1}}
\newcommand{\etabn}{\end{tabular}}
\newcommand{\beq}{\begin{equation}}
\newcommand{\eeq}{\end{equation}}
\newcommand{\beqn}{\begin{equation*}}
\newcommand{\eeqn}{\end{equation*}}
\newcommand{\bsplit}{\begin{split}}
\newcommand{\esplit}{\end{split}}

\newcommand{\gammadot}{\ensuremath{\dot{\gamma}}}
\newcommand{\Gsx}{\ensuremath{G_{\text{s}}(\Delta x, \Delta t)}}

\newcommand{\Gs}{\ensuremath{G_{\text{s}}}}

\newcommand{\kB}{\ensuremath{k_{\mathrm{B}}}}
\newcommand{\drna}{\ensuremath{d\bm{r}^{\mathrm{na}}}}

 \definecolor{green4}{rgb}{0,0.6,0}
 \definecolor{blue4}{rgb}{0,0,0.6}

\newcommand{\delete}[1]{}

\begin{document}

\title{Single particle fluctuations and directional correlations in driven hard sphere glasses}

\author{Suvendu Mandal}
\affiliation{Max-Planck Institut f\"ur Eisenforschung, Max-Planck Str.~1, 40237 D\"usseldorf, Germany}
\author{Vijaykumar Chikkadi}
\affiliation{Institute of Physics, University of Amsterdam, Science Park 904, 1098 XH Amsterdam, The Netherlands}
\author{Bernard Nienhuis}
\affiliation{Institute of Physics, University of Amsterdam, Science Park 904, 1098 XH Amsterdam, The Netherlands}
\author{Dierk Raabe}
\affiliation{Max-Planck Institut f\"ur Eisenforschung, Max-Planck Str.~1, 40237 D\"usseldorf, Germany}
\author{Peter Schall}
\affiliation{Institute of Physics, University of Amsterdam, Science Park 904, 1098 XH Amsterdam, The Netherlands}
\author{Fathollah Varnik}
\email{fathollah.varnik@rub.de}
\affiliation{Max-Planck Institut f\"ur Eisenforschung, Max-Planck Str.~1, 40237 D\"usseldorf, Germany}
\affiliation{Interdisciplinary Centre for Advanced Materials Simulation (ICAMS), Ruhr-Universit\"at Bochum, Stiepeler Strasse 129, 44801 Bochum, Germany}

\begin{abstract}
Via event driven molecular dynamics simulations and experiments, we study the packing fraction and shear-rate dependence of single particle fluctuations and dynamic correlations in hard sphere glasses under shear. At packing fractions above the glass transition, correlations increase as shear rate decreases: the exponential tail in the distribution of single particle jumps broadens and dynamic four-point correlations increase.  Interestingly, however, upon decreasing the packing fraction, a broadening of the exponential tail is also observed, while dynamic heterogeneity is shown to decrease. An explanation for this behavior is proposed in terms of a competition between shear and thermal fluctuations. Building upon our previous studies [Chikkadi et al, Europhys. Lett. (2012)], we further address the issue of anisotropy of the dynamic correlations.
\end{abstract}


\maketitle
\section{1. Introduction}

The relation between the dynamics of structural rearrangement and response in driven amorphous materials has been subject of growing interest in the past decade \cite{Argon1979,Varnik2003,Miyazaki2004,Lemaitre2009,Martens2011,Mandal2012}. A central issue in this context is the growth of spatial correlations upon approaching the glass transition (see, e.g., \cite{Bennemann1999a,Scheidler2002,Baschnagel2005,Ballesta2008,Varnik2009,Kob2012} and references therein). In quiescent glasses, it is well established that upon approaching the glass transition, dynamic correlations grow; this is manifested in an increase of the dynamic susceptibility~\cite{Berthier2011a} as well as strongly non-Gaussian displacement distributions. However, extensive simulations and experiments showed that besides the sudden increase of the macroscopic viscosity and relaxation time, the dynamic correlation length remains relatively small, limited to a few particle diameters, \cite{Scheidler2002,Ballesta2008,Varnik2009,Kob2012}. The situation changes when the glass is driven by applied shear that forces structural rearrangements~\cite{Chikkadi2011}; such external driving can lead to avalanche-like plastic response \cite{Lemaitre2009}. It is then interesting to elucidate how the dynamics crosses over from the thermal regime of supercooled liquids to the athermal limit of strongly driven glasses. Besides the magnitude of correlations, an important issue concerns their direction dependence: because shear introduces obvious directionality, this should reflect itself in the microscopic fluctuations or their correlations, setting it apart from thermally induced correlations. Indeed, our recent simulations and experiments revealed a transition from isotropic to anisotropic correlations with increasing importance of shear~\cite{Chikkadi2012b}; this observation is in qualitative agreement with the anisotropic correlations observed in the athermal limit of two-dimensional Lennard-Jones glasses~\cite{Maloney2009} and at finite temperatures~\cite{Furukawa2009}. These anisotropic correlations might play a central role in the shear-banding transition of glasses as suggested in~\cite{Chikkadi2011,Martens2011}.

Hard-sphere systems have been widely used as model systems to elucidate the dynamics of glasses because of their conceptual simplicity. These systems have been particularly valuable to obtain experimental insight into the microscopic dynamics. While the small molecular length scales make the direct observation of microscopic fluctuations in molecular glasses a difficult task, the situation is more favorable in the case of colloidal particles with diameters of the order of $1\mu$m thanks to the development of modern confocal microscopy techniques \cite{Besseling2007,Schall2007}. This technique has been fine-tuned in the recent years to allow simultaneous tracking of the trajectories of a large number of particles (up to $N=10^6$). We have recently combined this promising experimental tool with computer simulations to study the emergence of anisotropy of correlations in hard-sphere colloidal glasses~\cite{Chikkadi2012b}. A systematic investigation of the effect of shear versus packing fraction (the fraction of the system volume occupied by particles) on the microscopic fluctuations and their correlations, however, remained elusive; this would provide deeper insight into the nature of the dynamic arrest of glasses and the response to applied shear.

In this paper, we build upon our earlier studies and investigate the shear-rate and packing-fraction dependence of dynamic fluctuations and correlations in hard-sphere glasses. We present event-driven molecular dynamics simulations and experimental observations of particle dynamics in sheared supercooled liquids and glasses at shear-rates from the thermal to the shear-dominated regime.

We find that upon decreasing the shear rate, the exponential tail of the single-particle displacement distributions broadens, and we show that this broadening is linked to the increase of dynamic heterogeneity~\cite{Ediger2000,Berthier2011b} as proposed in \cite{Chaudhuri2007}. In contrast to this, upon increasing the packing fraction at constant shear rate, dynamic correlations grow but the exponential tail of the displacement distributions becomes narrower. We interpret this in terms of a competition between shear-induced and inherent thermally-induced dynamics. Finally, we address the question of the direction dependence of spatial correlations. We find a consistent
direction dependence in both simulations and experiments: isotropic correlations in the thermal regime go over into anisotropic correlations when shear dominates the relaxation. Besides this agreement, however, the decay of correlations has different functional form. While an exponential decay perfectly describes our simulation data, a power-law behavior is found in the experiments. This longer-range power-law decay is in qualitative agreement with the much larger correlation volume measured in the experiments. We hypothesize that hydrodynamic interactions, which are present in the experiments, but not accounted for in the simulations, might be responsible for the different behavior of correlations.

The paper is organized as follows: In section 2, we provide details of the simulations and the experimental system. We then present simulation results on the structure and single-particle displacements in section 3A, investigate the relation between single-particle displacements and dynamic correlations in section 3B, and elucidate the functional form of correlations in section 3C. Section 4 is devoted to the experimental results. We investigate the structure and single-particle displacements in section 4A, the dynamic correlations in section 4B, and elucidate the functional form in section 4C. Final discussions and conclusions are then presented in section 5.

\section{2. Simulation model and experimental system} \label{sec:simulation-setup}

Our simulation model is a polydisperse (11\%) hard sphere system of mass $m=1$ and average diameter $\sigma=0.8$. We perform event driven MD simulations \cite{DynamO}, and we do not observe any crystallization. Lengths are measured in units of the average particle diameter, $\sigma$. We use Lees-Edwards boundary conditions to have a time dependent shear strain $\gamma=t\dot \gamma$, where $\dot \gamma$ varies from $4\times10^{-5}$ to $10^{-2}$. With periodic boundary conditions on the coordinates $x_{i}$, $y_i$, and $z_{i}$ in an $L\times L\times L$ system, the position of particle $i$ in a box with strain $\gamma$ is defined as $r_{i}=(x_i+ \gamma z_i, y_i, z_i)$. The packing fraction is tuned from below to above the glass transition point, which, for the present polydisperse system, is located at a packing fraction of $\phi_{g}\approx 58.5\%$ \cite{Pussey2009}. The quiescent properties of this system have been studied extensively in \cite{Williams2001}. The temperature is fixed at $T=1$ via velocity rescaling. We present all the measurements after $100\%$ shearing to ensure that the system has reached the steady state.

For the experimental measurements, we use a hard-sphere suspension consisting of sterically stabilized polymethylmethacrylate (PMMA) particles suspended in a mixture of Cycloheptyl Bromide and Cis-Decalin that matches both the density and refractive index of the particles. The particles have a diameter of $\sigma = 1.3 \mu m$, with a polydisperity of $ 7\%$. The particle volume fraction is fixed at $\phi \sim 0.6$. We apply shear with constant shear rates in the range of $1.5 \times 10^{-5}$ to $2.2 \times 10^{-4} s^{-1}$, corresponding to Peclet numbers $\dot{\gamma} \tau$ between 0.3 and 2.2, respectively. Here, the structural relaxation time $\tau = 2 \times 10^4 s$ was determined from the mean-square displacement of the particles. More experimental details can be found in~\cite{Chikkadi2011}.

\section{3. Simulation Results}

\subsection{A. Structure versus single particle dynamics}

\begin{figure}
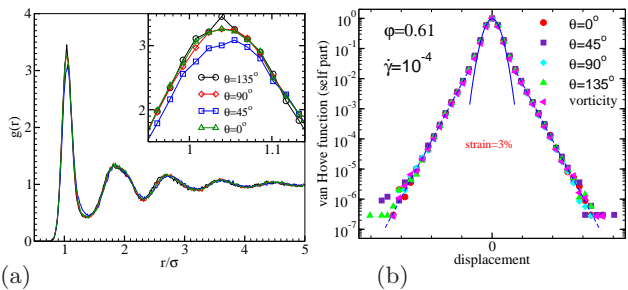

\unitlength=1mm
\begin{picture}(0,0)
\put(0,-2){(a)}
\put(50,-2){(b)}
\end{picture}
\includegraphics*[width=0.223\textwidth]{rdf_anisotropy_revised2.eps}
\includegraphics*[width=0.23\textwidth]{vanHove_function_for_different_angle_revised.eps}
\caption{(a) The pair distribution function along various directions. Interestingly, the difference between the principal coordinate directions ($\theta=0^\circ$ and $\theta=90^\circ$) is hardly detectable. Albeit small, the anisotropy is, however, well resolved when comparing the extension and compression directions ($\theta=45^\circ$ and $\theta=135^\circ$) in the shear plane (spanned by the flow and shear gradient directions).
(b) Distribution function of single particle displacements, $\Gsx$, determined along different directions as indicated. Obviously, no signature of anisotropy is visible in this quantity. Solid lines are exponential fits. The dashed line is a quasi-Gaussian fit.}
\label{fig:rdf+Gs}
\end{figure}

We first elucidate the glass structure and single-particle displacements in the simulations. From linear response theory it is well known that in a system under shear the pair distribution function cannot be fully isotropic, since it would then correspond to zero stress.  Some, even though slight, amount of anisotropy is thus always present in the structure of a sheared system. In order to highlight this property, we determine both via computer simulations and experiments, the pair distribution function, $g(r, \theta)$, along different spatial directions. Given that a particle is at the origin of the coordinate system, $g(r, \theta)$ is the probability density for finding another particle a distance $r$ apart from the origin along the direction $\theta$ with respect to the flow (in the shear plane).
Similar to previous experiments on colloidal suspensions \cite{Cheng2011} and computer simulations on a Lennard Jones glass \cite{Miyazaki2004}, we observe in \figref{fig:rdf+Gs}(a) an enhanced peak along the compression axis ($\theta=135^\circ$) and a weaker one along the extension axis ($\theta=45^\circ$). Furthermore, not quite unexpectedly, the peak position along the compression (extension) axis is shifted to slightly smaller (larger) distances. Interestingly, no detectable difference could be found for $g(r)$ along the $x$, $y$, and $z$ directions.

Since the local packing structure of an amorphous system strongly influences its dynamic behavior \cite{Gotze2009}, it is interesting to check whether this anisotropy has an effect on the distribution of single particle displacements. In this context, it must be mentioned that, in the presence of flow, the displacement of a particle consists of an affine part, which reflects the deterministic motion with the macroscopic velocity field, and a non-affine part, which constitutes the random part of the dynamics. Of course, when investigating the single particle distribution, it is the non-affine part of particle displacements which is used.

In order to obtain non-affine displacements, we proceed as follows \cite{Goldenberg2007}. For each particle, we follow nearest neighbors in time and determine the best affine tensor $\bm{\epsilon}$ that transforms the nearest neighbor vectors, ${\bf d}_i={\bf r}_i-{\bf r}_0$, over the time interval $\delta t$. This is done by minimizing $D^2 = (1/n) {\sum_{i=1}^{n}}({\bf d}_i(t + \delta t) - ({\bf I} + \bm{\epsilon})\cdot {\bf d}_i(t))^2$, where ${\bf I}$ is the identity matrix. $D^2$ reflects the mean-square deviation from a local affine deformation, and is an excellent  measure of local plasticity \cite{Falk1998}. The non-affine displacements of the particles are determined both (i) via subtraction of the local flow,

\begin{equation}
\drna={\bf r}(t+\delta t)- {\bf r}(t) - \int_0^t dt' {\bf u}(t', {\bf r}(t')),
\label{eq:drna}
\end{equation}
where ${\bf u}(t,{\bf r}(t))$ is a coarse-grained displacement field~\cite{Goldenberg2007} and (ii) as $\drna={\bf r}(t+\delta t)-{\bf r}(t) - \bm{\epsilon} \cdot  {\bf r}(t)$. It is assumed in (ii) that the coordinate center is at rest. We have verified that these two definitions give identical results as long as the shear is homogeneous across the channel~\cite{Chikkadi2012a}. In the general case of heterogeneous shear, however, the first definition is used.

Applying the above procedure, we have determined the distribution of non-affine displacements, also called the self part of the van Hove function, $\Gsx$, along various spatial directions. As seen from \figref{fig:rdf+Gs}(b), in contrast to the pair distribution, the single particle displacements seem to be quite insensitive to shear-induced anisotropy, in agreement with previous reports \cite{Varnik2007c}. In marked contrast to this observation, as will be shown below, the situation is quite different for the spatial correlations of the displacements.

\begin{figure}
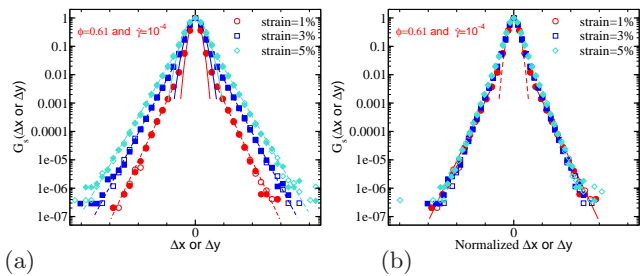

\unitlength=1mm
\begin{picture}(0,0)
\put(0,-2){(a)}
\put(50,-2){(b)}
\end{picture}
\includegraphics*[width=0.23\textwidth]{vanHove_function_for_different_strain.eps}
\includegraphics*[width=0.23\textwidth]{rescaled_vanHove_function_for_different_strain.eps}
\caption{(a) The self part of the van Hove function along the flow and vorticity directions ($x$ and $y$) for different strain intervals. (b) Same as in (a) but now normalized by the width of the distribution function.}
\label{fig:PDF_various_strain}
\end{figure}

\begin{figure}
\includegraphics*[width=0.4\textwidth]{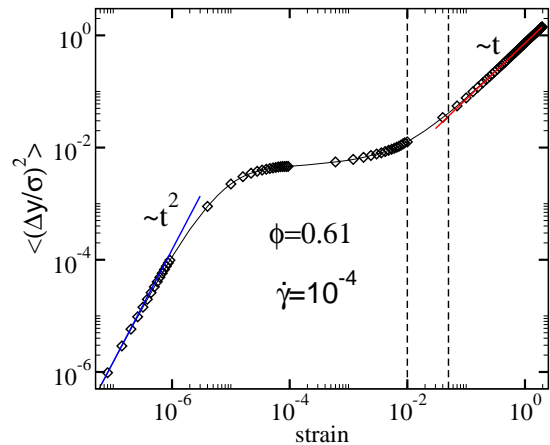}
\caption{Mean squared displacements at a packing fraction of $\phi=0.61$ and a shear rate of $\gammadot=10^{-4}$. The vertical dashed lines mark the range of strains used  in \figref{fig:PDF_various_strain}.}
\label{fig:msd}
\end{figure}

When computing the displacement distributions, a question arises regarding a possible dependence of the results on the selected time or strain interval. Here we show that there is a range of strain intervals where the shape of $\Gsx$ is essentially unchanged (\figref{fig:PDF_various_strain}). As a survey of the mean-square displacement (see Fig.~\ref{fig:msd}) clearly shows, the selected strain interval corresponds to the time domain, where particles start to leave the cage but still are partially trapped (departure from the plateau in the MSD). We expect significant changes in $\Gsx$ both for shorter and longer strain intervals. For shorter times, particles move essentially  unperturbed along straight lines (ballistic motion) and since the velocity distribution is a Maxwellian, the distribution of the displacements $\Delta {\bf r} = {\bf v} \Delta t$ is also a Maxwellian (Gaussian). In the limit of long times, on the other hand, particles leave the cage and their motion becomes uncorrelated to that of their neighbors so that again a Gaussian distribution is established \cite{Chaudhuri2007}.

\subsection{B. Single particle displacements and dynamic heterogeneity} \label{sec:structure}

\begin{figure}
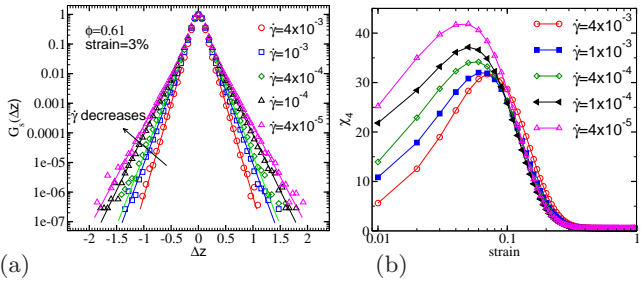

\unitlength=1mm
\begin{picture}(0,0)
\put(0,-2){(a)}
\put(50,-2){(b)}
\end{picture}
\includegraphics*[width=0.24\textwidth]{vanHove_function_for_different_shear_rate_revised.eps}
\includegraphics*[width=0.223\textwidth]{ki4_61_le_16k_revised.eps}
\caption{Effect of shear rate on (a) $G_{s}(\Delta x)$ and (b) $\chi_4$. The exponential tail becomes broader as $\gammadot$ decreases. This is accompanied by a corresponding increase of the maximum value of the dynamic susceptibility, a measure of dynamic heterogeneity in the system.}
\label{fig:Gsx+xi4-gammadot}
\end{figure}

We are now in the position to study the effect of shear rate and packing fraction on the displacement distribution. The effect of shear rate is demonstrated in \figref{fig:Gsx+xi4-gammadot}(a). We observe a Gaussian central part and a perfect exponential decay at large displacements; this exponential tail broadens with decreasing shear rate. Broad non-Gaussian tails in the self part of the van Hove function have been also observed in \cite{Chaudhuri2007}. The exponential form of the tail and its universal character has been recently addressed in \cite{Chaudhuri2007} where a simple   model based on the idea of dynamic heterogeneity has been proposed, which could reproduce this important feature with only a few fit parameters. The central idea behind the approach proposed in  \cite{Chaudhuri2007} is that the particles in the system can be divided into slow and fast groups. While the former perform essentially vibrational motion in the cage formed by their neighbors, the more mobile particles make comparably large jumps (of the order of the cage size). Despite the fact that the distribution of single particle displacements in both cases are Gaussian (with different underlying length and time scales), an  exponential tail with logarithmic corrections can be deduced for the $\Gsx$  of the entire system \cite{Chaudhuri2007}.

This motivates us to seek an interpretation of our observations in terms of dynamic heterogeneity. The observed broadening would imply that dynamic heterogeneity becomes enhanced at lower $\gammadot$. A way to test this idea is to compute the so-called correlation volume for dynamic correlations. There are a number of ways of determining the correlation volume. Two well-known possibilities are the overlap function \cite{Kob2012} and the peak value of the four-point susceptibility of density fluctuations, $\chi_4$ \cite{Berthier2011b}.  Here, we choose the second option and determine $\chi_4=N[\left< f^2_q(t) \right> - \left< f_q(t) \right>^2]$. In this expression, $N$ is the particle number and $f_q(t)=N^{-1}\sum^N_{i=1}\exp[ \mathrm{i} \mathbf{q} \cdot (\mathbf{r}_i(t) - \mathbf{r}_i(0) )]$ is the incoherent scattering function at wave vector $\mathbf{q}$. Results on $\chi_4$ are plotted in panel (b)
of \figref{fig:Gsx+xi4-gammadot}. In agreement with the above described picture, the maximum value of $\chi_4$ increases with decreasing shear rate. This observation is also in line with early studies of Yamamoto and Onuki who showed evidence for the growing length of dynamic correlations upon decreasing shear rate in a binary mixture of soft-core particles \cite{Yamamoto1998a}.

\begin{figure}
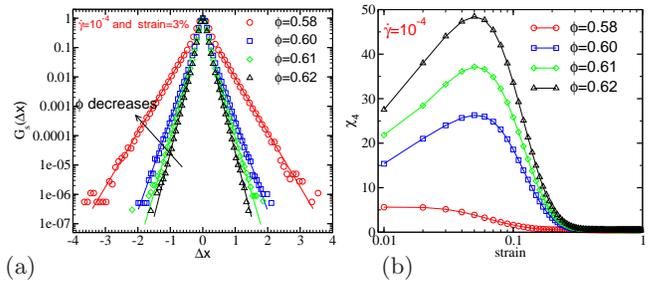

\unitlength=1mm
\begin{picture}(0,0)
\put(0,-2){(a)}
\put(50,-2){(b)}
\end{picture}
\includegraphics*[width=0.24\textwidth]{vanHove_function_for_different_volume_fraction_revised.eps}
\includegraphics*[width=0.223\textwidth]{ki4_for_diff_volume_fraction.eps}
\caption{Effect of packing fraction on (a) $G_{s}(\Delta x)$ and (b) $\chi_4$. The exponential tail in $\Gsx$ becomes broader as $\phi$ increases. This is accompanied by a growth of the maximum value of $\chi_4$.}
\label{fig:Gsx+xi4-phi}
\end{figure}

The effect of packing fraction is addressed in \figref{fig:Gsx+xi4-phi}. We show $\Gs$ and $\chi_4$ for a range of packing fractions at a fixed shear rate of $\gammadot=10^{-4}$. Again, a Gaussian central part and a perfect exponential decay at large displacements is found in all the cases shown. Note that, upon decreasing packing fraction, one expects a \emph{decrease} of correlations but an increase of particle mobility. We thus expect that the probability for a jump of a given length shall increase upon decreasing $\phi$, while the maximum of $\chi_4$ shall decrease. This expectation is confirmed by the data shown in \figref{fig:Gsx+xi4-phi}.

\subsection{C. Spatial correlations of displacements}

In this section, we focus on the direction-dependence of dynamic correlations. To do so, we use the above introduced scalar quantity $D^2$, which is actually a byproduct of the calculation of non-affine displacements. In order to study spatial correlation between non-affine displacements, we define the function~\cite{Chikkadi2011}
\begin{equation}
C_{D^2}({\Delta \mathbf{r}}) = \frac{ \left< D^2({\bf r} + \Delta \mathbf{r}) D^2(\mathbf{r})
\right> - \left< D^2(\mathbf{r}) \right> ^{2} } { \left< D^2(\mathbf{r})^{2}
\right> - \left< D^2(\mathbf{r}) \right> ^{2} }.
\label{c_r}
\end{equation}

The function $C_{D^2}(\Delta \mathbf{r})$ provides a measure of correlations between non-affine displacements at two points in space separated by a vector $\Delta {\mathbf r} =(\delta x, \delta y, \delta z$). Here, we determine directional dependence of this correlation function by projecting the distance vector $\Delta {\mathbf r}$ along different directions with respect to the flow. To avoid unnecessary fluctuations, we average the correlation over angular width $\pi/60$. It is also worth mentioning that, as shown above in the case of $\Gsx$, we have explicitly checked that the correlation function $C_{D^2}$ is insensitive to the specific value of the strain, $\delta \gamma$, as long as it belongs to the intermediate regime (data not shown).

\begin{figure}
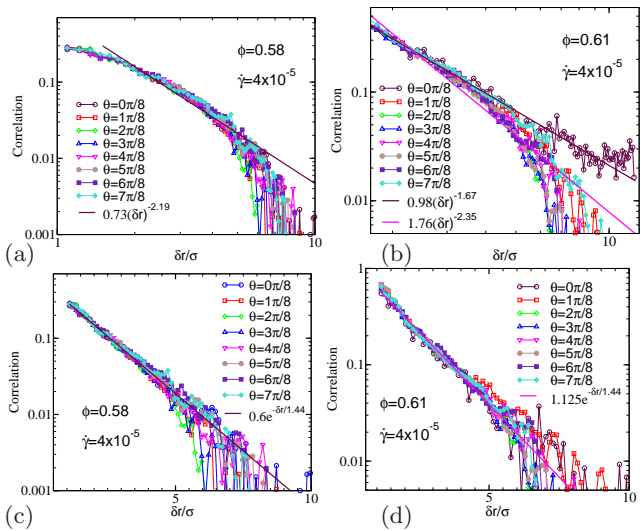

\unitlength=1mm
\begin{picture}(0,0)
\put(0,0){(a)}
\put(50,0){(b)}
\put(0,-35){(c)}
\put(50,-35){(d)}
\end{picture}
\includegraphics*[width=0.23\textwidth]{anisotropic_correlation_for_diff_angle_58_0.00004_power_law_fit.eps}
\includegraphics*[width=0.23\textwidth]{anisotropic_correlation_for_diff_angle_61_0.00004_power_law_fit.eps}
\includegraphics*[width=0.23\textwidth]{anisotropic_correlation_for_diff_angle_58_0.00004_exponential_fit2.eps}
\includegraphics*[width=0.23\textwidth]{anisotropic_correlation_for_diff_angle_61_0.00004_exponential_fit2.eps}
\caption[]{Correlation of plastic activity in the shear ($x,y$) plane along different spatial directions with respect to the flow. In order to better highlight the functional dependence, the same data are presented both in a log-log scale ((a) and (b)) and in a log-linear plot ((c) and (d)). In (d), the angle-dependent long time limit of $C_{D^2}(\Delta \mathbf{r})$ is subtracted from the data.}
\label{fig:CD2}
\end{figure}

Results on $C_{D^2}$ are depicted in \figref{fig:CD2} for two characteristic packing fractions of $\phi=0.58$ (supercooled state) and $\phi=0.61$ (glass) for a shear rate of $\gammadot=4 \times 10^{-5}$. As seen from this plot, the correlation of plastic activity is fully isotropic in the supercooled state, while marked anisotropy is visible in the glassy phase ($\phi=61$). This is an important observation since it tells us that, even in the steady state where a colloidal glass is considered to be completely shear melted, it is possible to distinguish a glass from a supercooled liquid by purely dynamical measurements, i.e., without referring to any static property of the system.

In order to highlight the functional form of $C_{D^2}({\Delta \mathbf{r}})$, we have illustrated the above data both in log-linear and in log-log scale. As visible from panels (a) and (b), strong deviations from the power law are observed both in the supercooled state and in the glass. This is in line with our previous report at a slightly higher shear rate \cite{Chikkadi2012b}. Interestingly, plotting the same data in the log-linear scale reveals perfect exponential decay both in the supercooled state (\figref{fig:CD2}(c)) and in the glassy phase (\figref{fig:CD2}(d)). We are not aware of any theoretical explanation for this exponential decay, and will show below that this in fact contrasts with the experimental observations. A possible interpretation of the different behavior of correlations in simulations and experiments is also given below.

\section{4. Experimental results}

\subsection{A. Structure versus single particle displacements}

We complement our simulations with experimental measurements of particle displacements in colloidal glasses. We first elucidate changes in the glass structure under shear by
\begin{figure}
\unitlength=1mm
\begin{picture}(0,0)
\put(0,-2){(a)}
\put(50,-2){(b)}
\end{picture}
\includegraphics*[width=0.46\textwidth]{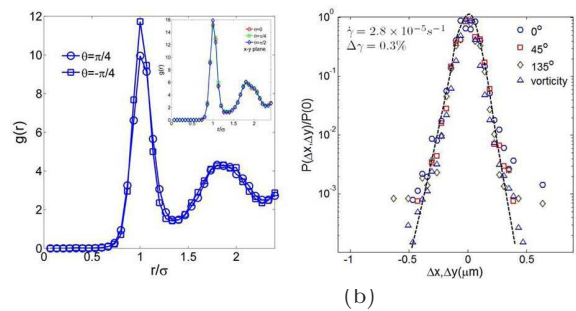}
\caption{Experimental pair distribution function (a) and particle displacements (b) at a shear rate of $\dot{\gamma} \tau = 2.2$. The main panel in (a) shows the pair distribution function along the extension ($-\pi/4$) and compression direction ($\pi/4$) in the shear plane. A clear difference is observed. Inset shows the pair distribution function in the flow-vorticity plane. No significant difference between the directions is observed. (b) Particle displacements in the shear plane along the indicated directions with respect to the flow axis, as well as along the vorticity direction. Good overlap is observed.}
\label{Exp:g(r)}
\end{figure}
showing experimentally measured pair distribution functions in Fig.~\ref{Exp:g(r)}a. Similar to the simulations, in Fig.~\ref{Exp:g(r)}a, main panel, we observe an enhanced peak along the compression direction, and a weaker one along the dilation direction, demonstrating a small distortion of the structure in the shear plane. On the other hand, no significant difference is observed in the flow - vorticity plane as shown in the inset, again in agreement with the simulations. Note that because of the lower resolution of the confocal microscope in the vertical (flow gradient) direction, the absolute peak values in the main panel and inset are different. This is also the reason why we cannot directly compare directions with different inclinations to the vertical (flow gradient) direction. Nevertheless, comparison of directions with the same inclination to the vertical axis such as those shown in Fig.~\ref{Exp:g(r)}a is meaningful, and indeed show a small structural distortion similar to the simulations.

\begin{figure}
\unitlength=1mm
\includegraphics*[width=0.48\textwidth]{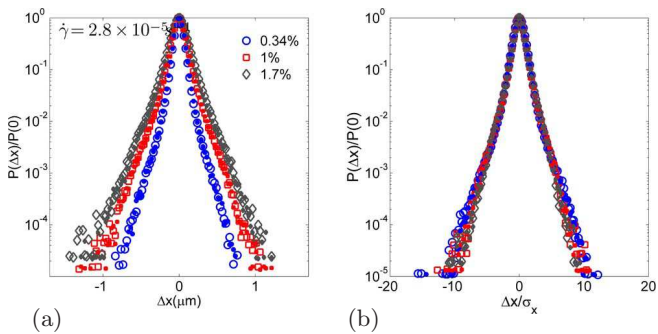}
\begin{picture}(0,0)
\put(-40,2){(a)}
\put(2,2){(b)}
\end{picture}
\caption{Experimentally measured particle displacements for increasing strain intervals. (a) Probability of particle displacements along the flow (open symbols) and vorticity direction (closed symbols). (b) Same as in (a), but now normalized by the width of the distribution. The robust shape of the distribution function indicates robust behavior of the glass in the investigated time domain.}
\label{fig:exp_displacements}
\end{figure}

Despite the small anisotropy of the structure, the displacements of the particles are surprisingly isotropic, again in agreement with the simulations. To show this, we determine the non-affine displacements by subtracting contributions from the mean flow according to eq.~(\ref{eq:drna}). We show the non-affine displacements resolved along various directions with respect to the flow direction in Fig.~\ref{Exp:g(r)}b. The data shows that the distributions overlap, indicating that the single particle displacements are isotropic. This is further confirmed when we investigate displacements for different strain intervals; the corresponding distributions of displacements, resolved along the flow and vorticity directions, are indicated in Fig.~\ref{fig:exp_displacements}. The data shows that the two distributions overlap for all strain intervals, again indicating the isotropy of the single particle displacements. In fact, similar to the simulations, the displacement distributions remain robust for all investigated strain intervals as shown by the overlap of the rescaled curves in Fig.~\ref{fig:exp_displacements}b. The collapse of the data indicates that the strain intervals all probe a similar time domain, where the particles exhibit essentially similar diffusive characteristics. The mean-square displacement of the particles (data not shown) shows that this time domain corresponds to the onset of the diffusive regime, similar to the regime addressed by the simulations, as shown in Fig.~\ref{fig:msd}.

\subsection{B. Single-particle displacements and dynamic heterogeneity}

\begin{figure}
\unitlength=1mm
\begin{picture}(0,0)
\put(0,-2){(a)}
\put(50,-2){(b)}
\end{picture}
\includegraphics*[width=0.23\textwidth]{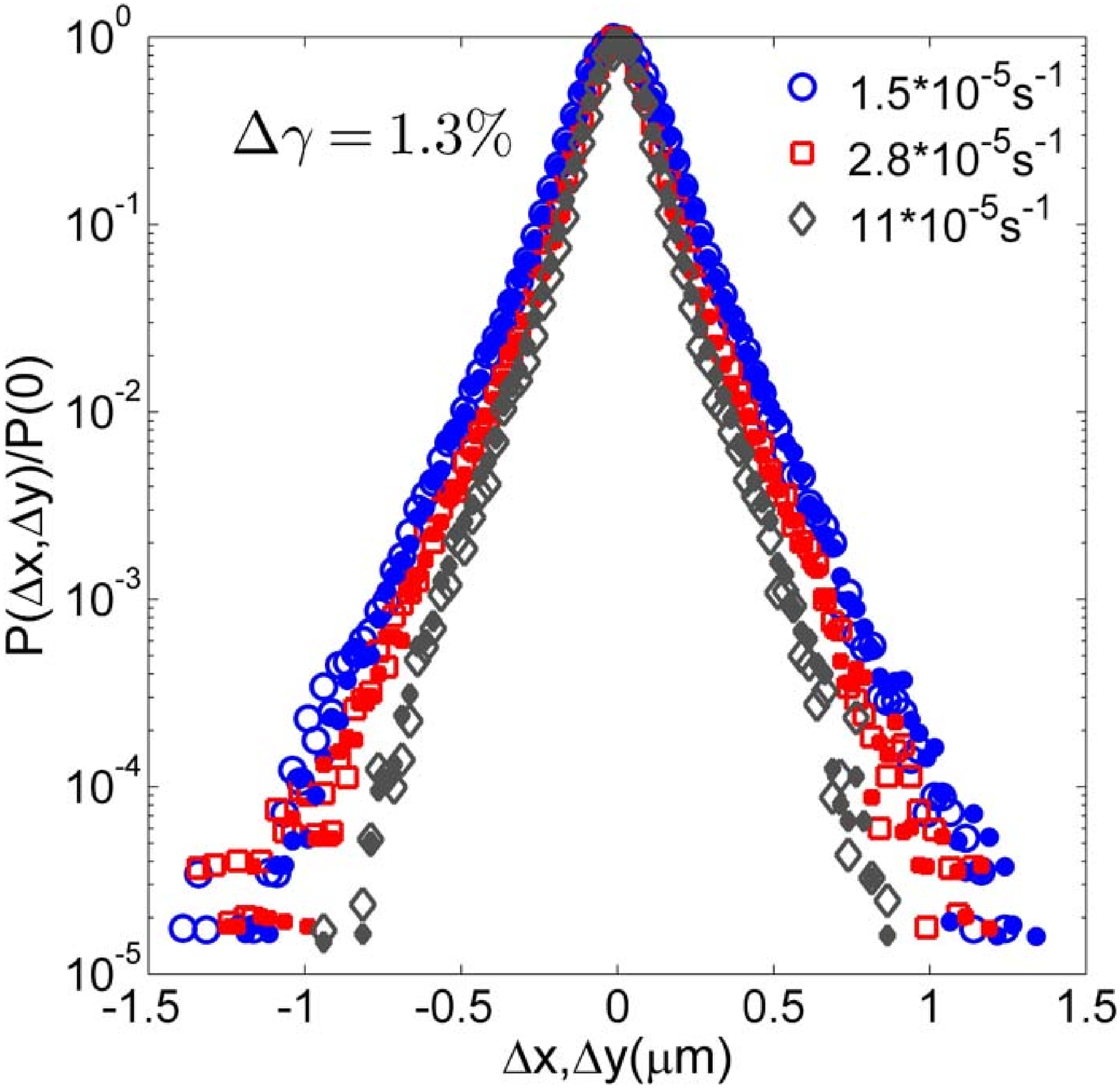}
\includegraphics*[width=0.233\textwidth]{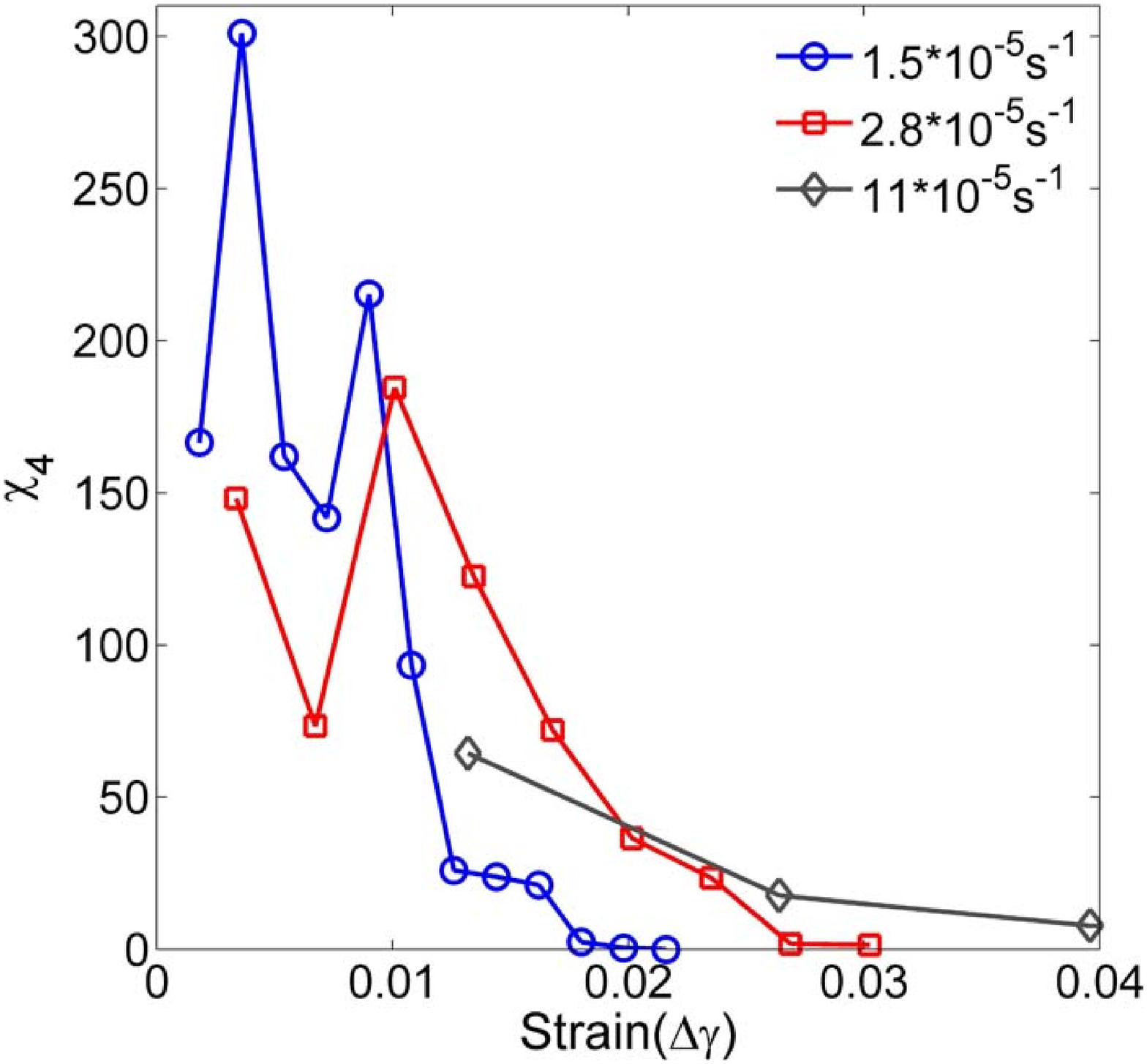}
\caption{Experimentally measured displacement distribution (a) and dynamic susceptibility (b) plotted for increasing shear rate. The exponential tail in (a) becomes broader as $\dot{\gamma}$ decreases, similar to the simulations. This is accompanied by an increase of the dynamic susceptibility as shown in (b).}
\label{fig:chi4}
\end{figure}

Similar to the simulations, we use the full time-dependent particle trajectories to determine the corresponding displacement distributions and their dynamic correlations. We plot the results for three different shear rates spanning Peclet numbers from 0.3 to 2.2 in Fig.~\ref{fig:chi4}. In agreement with the simulations, the displacement distributions show a broadening of the exponential tail with decreasing strain rate indicating increasing dynamic heterogeneity (Fig~\ref{fig:chi4}a). This increasing dynamic heterogeneity is confirmed by measurement of the dynamic susceptibility as shown in Fig.~\ref{fig:chi4}b. At the smallest applied strain rate, the dynamic susceptibility rises to values larger than $\sim 300$, indicating long-range correlated motion. Due to the limited acquisition speed of the 3D imaging, however, small strain intervals are not accessible for the fastest strain rate and the maximum dynamic susceptibility lies outside the accessible window. While the results in Fig.~\ref{fig:chi4} are in qualitative agreement with the simulations shown in Fig.~\ref{fig:Gsx+xi4-gammadot}, the absolute value of the dynamic susceptibility observed here is much larger than in the simulations. Obviously, this discrepancy cannot be explained by a packing fraction effect as the overview over different packing fractions in Fig. ~\ref{fig:Gsx+xi4-phi}b shows. A possible origin of this discrepancy is the hydrodynamic interaction between the particles that can have an influence on the collective dynamics at intermediate times, where the particles are rattling in the cage formed by their neighbors.

We are aware of the fact that the externally imposed shear rates studied in our experiments are far too low to give rise to macroscopically measurable lubrication forces so that the rheology of the system is largely independent of lubrication interactions. Regarding spatial correlations of particle displacements, however, the following argument may nevertheless be relevant. A colloidal particle of one micron diameter has a thermal velocity of the order of
$v_{\text{thermal}} \sim \sqrt{\kB T/m_{\text{Colloid}}} \sim 10^{-3}$m/s. On the other hand, at high packing fractions considered in our studies, the surface-to-surface distance of two neighboring colloids is of the order of 10\% of their diameter, i.e., $d_{12} \sim 10^{-7}$m. A thermal motion of a colloidal particle with respect to its neighbor, thus gives rise to a local shear rate of the order of $\gammadot \sim v_{\text{thermal}}/d_{12}\sim 10^4$/s, many orders of magnitude higher than the applied shear rate. The resulting hydrodynamic interactions can give rise to longer-range correlations: The neighboring particle will be subject to a short living but relatively high shear stress of the order $\sim \eta_0 \gammadot$, with $\eta_0$ being the viscosity of the bare solvent. Thus, if a particle moves via its thermal motion along a given direction, it may give rise to a similar motion of its neighboring particle along the same direction. The spatial correlations produced by this mechanism are expected to occur along the direction transverse to the motion of particle. Another hydrodynamic effect, giving rise to longitudinal correlations, would arise from the incompressibility of the solvent. As a colloidal particle performs a jump along a given direction, the particle behind it will tend to follow along the same direction due to the solvent flow. In our MD simulations, however, there is no solvent so that all these effect are absent.

\section{C. Spatial correlations}
\begin{figure}
\unitlength=1mm
\includegraphics*[width=0.5\textwidth]{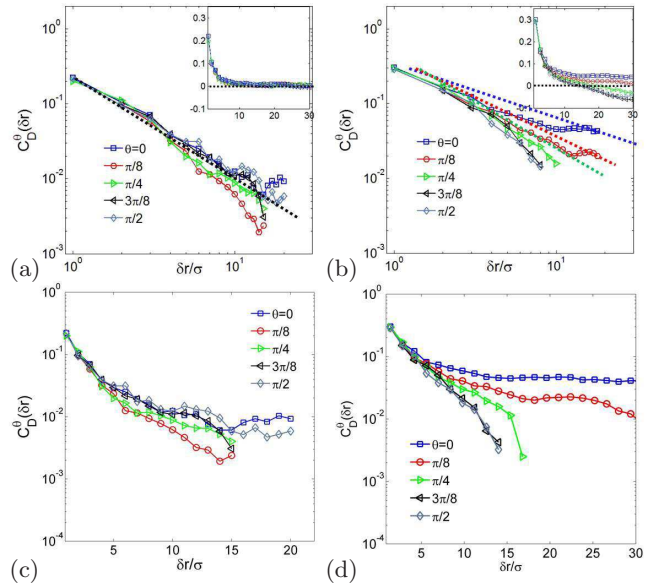}
\begin{picture}(0,0)
\put(-42,45){(a)}
\put(0,45){(b)}
\put(-42,5){(c)}
\put(0,5){(d)}
\end{picture}
\caption{Experimental correlation functions of non-affine displacements in the thermal ($\dot{\gamma} \tau = 0.3$) (a,c) and shear-dominated regime ($\dot{\gamma} \tau = 2.2$) (b,d), in double logarithmic (upper row) and half-logarithmic representation (lower row). The correlation functions are resolved along the indicated directions with respect to the flow direction in the shear plane. }
\label{fig:correlations}
\end{figure}

Finally, we elucidate the direction-dependence of the dynamic correlations. We do so by determining the full three-dimensional correlation function from the measured particle trajectories, as done previously in the simulations. Again, we use the non-affine part of the displacements (eq.~(\ref{c_r})) to be not affected by contributions from the mean flow. To smooth the experimental data, this time we choose angular bins of width $\pi / 18$ around the specific directions. The resulting correlation functions are shown in Fig.~\ref{fig:correlations}. We distinguish two regimes: the thermal regime, where $\dot{\gamma} \tau < 1$ and particle motion is dominated by thermal fluctuations (Fig.~\ref{fig:correlations}a,c), and the regime $\dot{\gamma} \tau > 1$, where particle displacements are dominated by the applied shear (Fig.~\ref{fig:correlations}b,d). The data reveals a characteristic change of the decay of correlations similar to that observed in the simulations: correlations exhibit isotropic decay in the thermal, and anisotropic decay in the shear-dominated regime. In the latter case, the anisotropic decay is characterized by a slower decay in the flow- and a faster decay in the flow-gradient direction. In contrast to the simulations, however, the functional form of the decay appears different. The experimental data in Fig.~\ref{fig:correlations} suggest a power-law decay of correlations, whereas the simulations indicate an exponential decay. This slower decay is in qualitative agreement with the larger values of $\chi_4$ observed before: larger values of $\chi_4$ indicate that dynamic correlations span more particles, and are therefore more extended in space, in agreement with the power-law decay observed in the experiments. As mentioned above, a possible reason for these longer-ranged spatial correlations may be solvent-mediated hydrodynamic interactions, which are present in experiments but not accounted for in our event driven MD simulations.

\section{5. Conclusion}

Our simulations and experiments reveal interesting properties of displacement fluctuations in sheared glasses. While the glass structure and single-particle displacement fluctuations remain essentially isotropic, interesting features arise in the correlations of these displacements in terms of their anisotropy, volume-fraction and strain-rate dependence. Our detailed analysis of simulation and experimental data shows that within the range of shear rates studied here, dynamic correlations grow with decreasing applied strain rate. This is mirrored in an increasing non-Gaussian behavior of the displacement distributions. Furthermore, correlations grow with increasing packing fraction. An interesting transition arises with respect to the symmetry of correlations: while correlations decay isotropically in the thermal regime, they become anisotropic when shear dominates the displacements. We find overall good qualitative agreement between the simulations and experiments; however, the range of dynamic correlations differs significantly as evidenced by a short-range exponential decay of correlations in the simulations, and a longer-range power-law-like decay in the experiments. A possible reason for this discrepancy is given in terms of local hydrodynamic interactions which are present in the colloidal system but fully absent in simulations. Further work is needed to investigate the origin of this discrepancy more closely. Since this interpretation is based on hydrodynamic effects arising from short living cage rattling motion of particles, temperature may also play a role since it directly influences this type of particle motion. Finally, we note that, within the intermediate time domain considered, our observations are robust. However, outside this time domain, in the ballistic regime of much smaller and the diffusive regime of much longer time scales, correlations should become short-ranged again; this behavior should thus be similar to the behavior of dynamic correlations in quiescent glasses.

\section{acknowledgments}
S.M. is financially supported by the Max-Planck Society. Simulations are performed using DynamO \cite{DynamO}. P.S. acknowledges support by a VIDI fellowship from the Netherlands Organization for Scientific Research (NWO). ICAMS acknowledges funding from its industrial sponsors, the state of North-Rhine Westphalia and the European Commission in the framework of the European Regional Development Fund (ERDF).
\bibliographystyle{prsty_with_title}
\bibliography{literature_suvendu_mandal_v3}

\begin{thebibliography}{10}

\bibitem{Argon1979}
A.~S. Argon, Plastic deformation in metallic glasses, Acta Matrial {\bf 27},
  47  (1979).

\bibitem{Varnik2003}
F. Varnik, L. Bocquet, J.-L. Barrat, and L. Berthier, Shear localization in a
  model glass, Phys. Rev. Lett. {\bf 90},  095702  (2003).

\bibitem{Miyazaki2004}
K. Miyazaki, D. Reichman, and R. Yamamoto, Supercooled Liquids Under Shear:
  Theory and Simulation, Phys. Rev. E {\bf 70},  011501  (2004).

\bibitem{Lemaitre2009}
A. Lemaitre and C. Caroli, Rate-Dependent Avalanche Size in Athermally Sheared
  Amorphous Solids, Phys. Rev. Lett. {\bf 103},  065501  (2009).

\bibitem{Martens2011}
K. Martens, L. Bocquet, and J.-L. Barrat, Connecting Diffusion and Dynamical
  Heterogeneities in Actively Deformed Amorphous Systems, Phys. Rev. Lett. {\bf
  106},  156001  (2011).

\bibitem{Mandal2012}
S. Mandal, M. Gross, D. Raabe, and F. Varnik, Heterogeneous Shear in Hard
  Sphere Glasses, Phys. Rev. Lett. {\bf 108},  098301  (2012).

\bibitem{Bennemann1999a}
C. Bennemann, C. Donati, J. Baschnagel, and S.~C. Glotzer, Growing range of
  correlated motion in a polymer melt on cooling towards the glass transition,
  Nature {\bf 399},  246  (1999).

\bibitem{Scheidler2002}
P. Scheidler, W. Kob, and K. Binder, Cooperative motion and growing length
  scales in supercooled confined liquids, Europhys. Lett. {\bf 59},  701
  (2002).

\bibitem{Baschnagel2005}
J. Baschnagel and F. Varnik, Computer simulation of supercooled polymer melts
  in the bulk and in confined geometry, J.Phys.: Condens. Matter {\bf 17},
  R851  (2005).

\bibitem{Ballesta2008}
P. Ballesta, A. Duri, and L. Cipelletti, Unexpected drop of dynamical
  heterogeneities in colloidal suspensions approaching the jamming transition,
  Nature Phys. {\bf 4},  550  (2008).

\bibitem{Varnik2009}
F. Varnik and K. Binder, Multiscale modelling of polymers at interfaces, Int.
  J. Mater. Res. {\bf 100},  1494  (2009).

\bibitem{Kob2012}
W. Kob, S. {Roldan-Vargas}, and L. Berthier, Non-monotonic temperature
  evolution of dynamic correlations in glass-forming liquids, Nature Phys. {\bf
  8},  164  (2012).

\bibitem{Berthier2011a}
L. Berthier, Dynamic Heterogeneity in Amorphous Materials, Physics {\bf 4},  42
   (2011).

\bibitem{Chikkadi2011}
V. Chikkadi, G. Wegdam, D. Bonn, B. Nienhuis, and P. Schall, Long-Range Strain
  Correlations in Sheared Colloidal Glasses, Phys. Rev. Lett. {\bf 107},
  198303  (2011).

\bibitem{Chikkadi2012b}
V. Chikkadi, S. Mandal, B. Nienhuis, D. Raabe, F. Varnik, and P. Schall,
  Shear-induced anisotropic decay of correlations in hard-sphere colloidal
  glasses, Europhys. Lett. {\bf 100},  56001  (2012).

\bibitem{Maloney2009}
C.~E. Maloney and M.~O. Robbins, Anisotropic Power Law Strain Correlations in
  Sheared Amorphous 2D Solids, Phys. Rev. Lett. {\bf 102},  225502  (2009).

\bibitem{Furukawa2009}
A. Furukawa, K. Kim, S. Saito, and H. Tanaka, Anisotropic Cooperative
  Structural Rearrangements in Sheared Supercooled Liquids, Phys. Rev. Lett.
  {\bf 102},  016001  (2009).

\bibitem{Besseling2007}
R. Besseling, E.~R. Weeks, A.~B. Schofield, and W.~C.~K. Poon,
  Three-Dimensional Imaging of Colloidal Glasses under Steady Shear, Phys. Rev.
  Lett. {\bf 99},  028301  (2007).

\bibitem{Schall2007}
P. Schall, D.~A. Weitz, and F. Spaepen, Structural Rearrangements That Govern
  Flow in Colloidal Glasses, Science {\bf 318},  1895  (2007).

\bibitem{Ediger2000}
M.~D. Ediger, , Ann. Rev. Phys. Chem. {\bf 51},  99  (2000).

\bibitem{Berthier2011b}
L. Berthier, G. Biroli, J.-P. Bouchaud, L. Cipeletti, and W.~v. Saarloos, {\em
  Dynamical heterogeneities in glasses, colloids, and granular media} (Oxford
  University Press, Oxford, 2011).

\bibitem{Chaudhuri2007}
P. Chaudhuri, L. Berthier, and W. Kob, Universal Nature of Particle
  Displacements close to Glass and Jamming Transitions, Phys. Rev. Lett. {\bf
  99},  060604  (2007).

\bibitem{DynamO}
M. Bannerman, R. Sargant, and L. Lue, DynamO: a free equation O(N) general
  event-driven molecular dynamics simulator, J. Computatational Chem. {\bf 32},
   3329  (2011).

\bibitem{Pussey2009}
P.~N. Pussey, E. Zaccarelli, C. Valeriani, E. Sanz, W.~C.~K. Poon, and M.~E.
  Cates, Hard spheres: crystallization and glass formation, Phil. Trans. R.
  Soc. A {\bf 367},  4993  (2009).

\bibitem{Williams2001}
S.~R. Williams, I.~K. Snook, and W. van Megen, Molecular dynamics study of the
  stability of the hard sphere glass, Phys. Rev. E {\bf 64},  021506  (2001).

\bibitem{Cheng2011}
X. Cheng, H. McCoy, J.~N. Israelachvili, and I. Cohen, Imaging the microscopic
  structure of shear thinning and thickening colloidal suspensions, Science
  {\bf 333},  1276  (2011).

\bibitem{Gotze2009}
W. G\"otze, {\em Complex Dynamics of Glass-Forming Liquids-A Mode-Coupling
  Theory} (Oxford University, Oxford, 2009).

\bibitem{Goldenberg2007}
C. Goldenberg, A. Tanguy, and J.~L. Barrat, Particle displacements in the
  elastic deformation of amorphous materials: Local fluctuations vs. non-affine
  field, Europhysics Letters {\bf 80},  16003  (2007).

\bibitem{Falk1998}
M.~L. Falk and J.~S. Langer, Dynamics of viscoplastic deformation in amorphous
  solids, Phys. Rev. E {\bf 57},  7192  (1998).

\bibitem{Chikkadi2012a}
V. Chikkadi and P. Schall, Nonaffine measures of particle displacements in
  sheared colloidal glasses, Phys. Rev. E {\bf 85},  031402  (2012).

\bibitem{Varnik2007c}
F. Varnik,  in {\em (5th International Workshop on Complex Systems, 25-28
  September 2007, Sendai, Japan)} (American Institute of Physics, ADDRESS,
  2007), Vol.~CP 982, p.\ 160.

\bibitem{Yamamoto1998a}
R. Yamamoto and A. Onuki, Dynamics of highly supercooled liquids:
  Heterogeneity, rheology and diffusion, Phys. Rev. E {\bf 58},  3515  (1998).

\end{thebibliography}
\end{document}